\documentclass{article}
\usepackage{spconf,amsmath,graphicx}
\usepackage{amssymb}
\usepackage[table]{xcolor}
\usepackage{color}
\usepackage{balance}
\usepackage{multirow}

\usepackage[]{algorithm2e}
\usepackage{wasysym}
\usepackage{bm}
\usepackage{url,cite}
\usepackage{enumitem} 
\usepackage{pifont}

\def\tthline{\noalign{\hrule height 1.4pt}}
\def\thline{\noalign{\hrule height 1.0pt}}

\newcolumntype{?}{!{\vrule width 1pt}}

\title{Simultaneous Separation and Transcription of Mixtures \\ with Multiple Polyphonic and Percussive Instruments}

\name{Ethan Manilow, Prem Seetharaman, Bryan Pardo\thanks{This work has made use of the Mystic (Programmable Systems Research Testbed to Explore a Stack-WIde Adaptive System fabriC) NSF-funded infrastructure at Illinois Institute of Technology, NSF award CRI-1730689.
.}}
\address{Interactive Audio Lab\\
Northwestern University\\
Evanston, IL, USA}

\begin{document}

\maketitle

\begin{abstract}
We present a single deep learning architecture that can both separate an audio recording of a musical mixture into constituent single-instrument recordings \textit{and} transcribe these instruments into a human-readable format at the same time, learning a shared musical representation for both tasks. This novel architecture, which we call Cerberus, builds on the Chimera network for source separation by adding a third ``head'' for transcription. By training each head with different losses, we are able to jointly learn how to separate and transcribe up to five instruments with a single network. We show that separation and transcription are highly complementary with one another and when learned jointly, lead to Cerberus networks that are better at both separation and transcription and generalize better to unseen mixtures.
\end{abstract}

\begin{keywords}
source separation, music transcription, multitask learning, deep clustering, computer audition
\end{keywords}

By listening carefully to a musical mixture, humans are not only capable of attending to different sources in the mixture (e.g. focusing on the violin in a string quartet) but also of converting the activity of that source into a musical score (e.g. writing down the notes the violin is playing). Inspired by this, we have developed a system to simultaneously separate polyphonic instruments from a mixture of polyphonic instruments and also produce a piano roll transcription for each separated instrument. The core of our system is Cerberus\footnote{
https://interactiveaudiolab.github.io/demos/cerberus}, a novel deep learning architecture capable of learning to separate and transcribe polyphonic and percussive sources in complex musical mixtures. 

\begin{figure}
    \centering
    \includegraphics[width=1.0\linewidth]{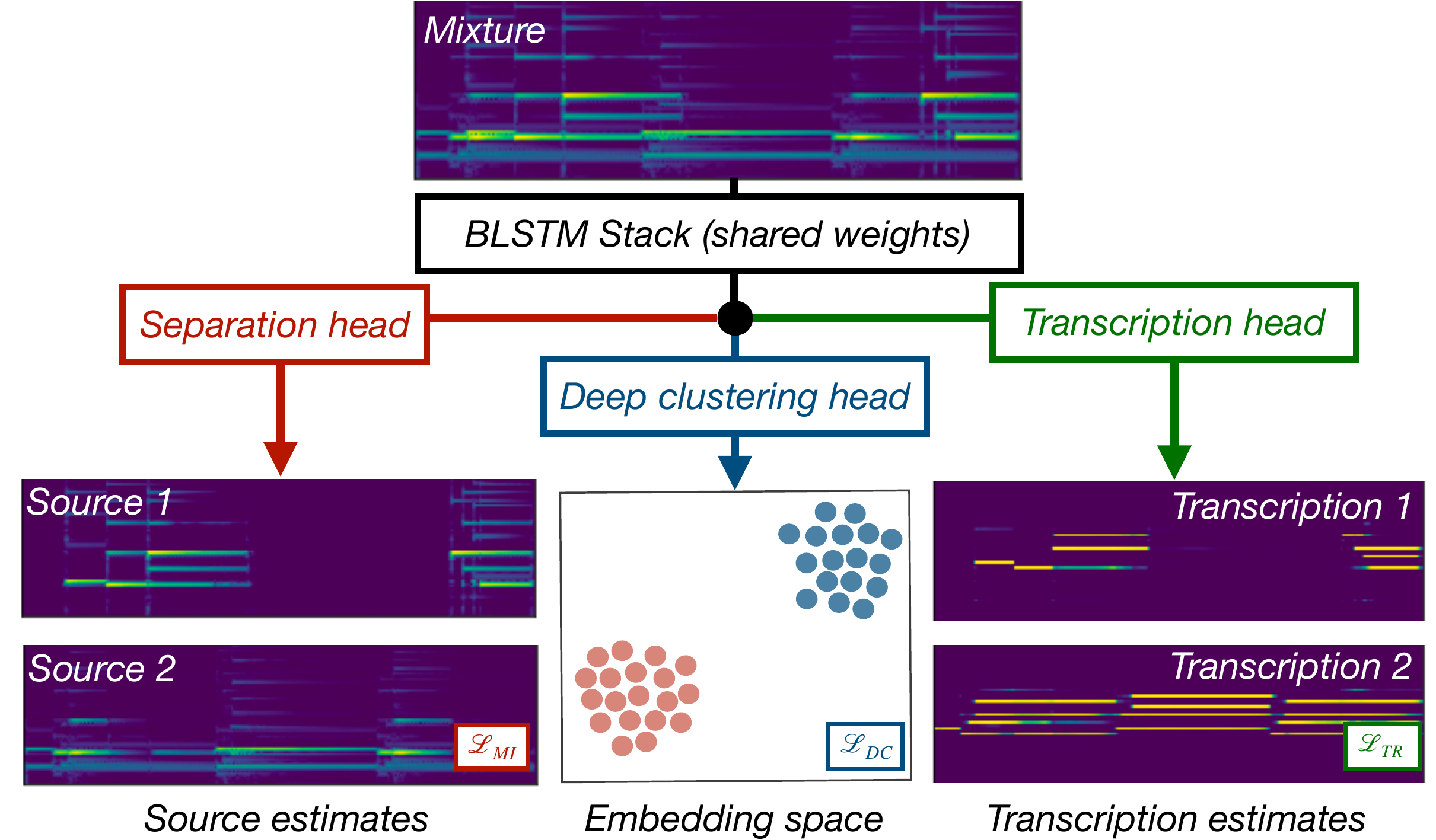}
    \caption{System overview of the Cerberus architecture. The input is the magnitude spectrogram of a musical mixture. There are three outputs (heads): an embedding space (trained via $\mathcal{L}_{DC}$ - deep clustering loss), estimated sources (trained via $\mathcal{L}_{MI}$ - mask inference loss) and the piano roll transcription of each source (trained via $\mathcal{L}_{TR}$ - transcription loss).}
    \label{fig:schematic}
    \vspace{-0.1cm}
\end{figure}

Audio source separation is the task of extracting a single source (e.g. a guitar) from an auditory mixture with multiple overlapping sources (e.g. a rock band). Source separation has advanced greatly recently, with many neural network architectures achieving impressive results for the task of musical source separation~\cite{hershey2016deep, erdogan2015phase, luo2017deep}. Most of these systems are trained only to separate audio sources. The output of these systems can enable downstream tasks \cite{chang2019mimo}. Here, instead of only separating sources or using the separation to enable a downstream task, we instead jointly learn two tasks: automatic music transcription and audio source separation. In our system, the two tasks are able to leverage a shared learned representation. 

The goal of Automatic Music Transcription (AMT) is to transform raw audio signals of music into a symbolic musical score (\textit{e.g.,} piano roll). AMT enables many other tasks in music information retrieval, such as query-by-humming \cite{birmingham2006query}, chord recognition \cite{ryynanen2008automatic} and computational musicology \cite{benetos2013automatic}. AMT systems are generally designed to transcribe a single monophonic (e.g. clarinet) or polyphonic (e.g. piano) source into a musical score \cite{benetos2013automatic}. Converting percussion recordings into notation is often regarded as a separate task \cite{cartwright2018increasing}. Many musical recordings contain multiple percussive and harmonic instruments sounding simultaneously. We address transcribing multiple simultaneous instruments (both percussive and polyphonic) into a individual piano roll per instrument.

Score informed music source separation~\cite{ewert2014score, duan2011soundprism} is related to our work. Here, musical score information guides source separation. These approaches require the score as input, whereas we use the score only as a training objective. In use, our system does not require a score, as it produces its own transcript. 

Our work fits more broadly into the field of multitask learning. The goal of multitask learning is to leverage commonalities between related tasks in an effort to generalize better on those tasks~\cite{ruder2017overview}. In audio, multitask learning has been leveraged in some related work. Hung \textit{et. al.,}~\cite{hung2019multitask} present a system that does frame-level instrument detection and pitch prediction. 
Existing work has to combined separation with related tasks, like F0 estimation \cite{jansson2019joint}.
We focus on learning jointly with separation the harder task of automatic music transcription of polyphonic rather than monophonic instruments as applied to arbitrary musical instruments, not just vocals.

\vspace{-0.10cm}
\section{Proposed method}
\vspace{-0.10cm}
Our proposed method simultaneously separates and transcribes musical mixtures using a deep net called Cerberus (depicted in Fig \ref{fig:schematic}), a ``three-headed'' network, where each head has a different output and a different objective function. The key idea is to first transform the input representation (e.g. a spectrogram) via shared processing layers into a learned representation useful for both transcription and separation. The learned representation can then be processed by smaller networks that are specialized for separation or transcription.

\vspace{-0.10cm}
\subsection{Source separation}
\vspace{-0.10cm}

Assume an audio mixture is in a time-frequency representation, such as a short term Fourier transform (STFT), represented by a matrix $\mathbf{X}$. Here, each element $\mathbf{X}(t,f)$ indicates the magnitude and phase of the mixture at time $t$ and frequency $f$. We perform a variant of mask-based source separation, where the goal is to make a non-negative mask matrix $\mathbf{M}_k$ for each sound source $k$, with values normalized to the interval $[0,1]$. Each element $\mathbf{M}_k(t,f)$ indicates the degree to which the energy in the auditory mixture at time $t$ and frequency $f$ is due to source $k$. To isolate source $k$, one can element-wise multiply the magnitude of the STFT $\mathbf{|X|}$ by the mask: $\mathbf{M}_k \circ \mathbf{|X|}$.

To train a deep net to provide output useful for mask-based source separation, ground truth training mixtures and their corresponding ideal source masks are provided and a loss function is used to measure the difference between network output and the ideal output. We build on two similar prior works: deep clustering \cite{hershey2016deep} and the Chimera architecture \cite{luo2017deep}. 

In deep clustering, a neural network is trained to map each time-frequency bin in $|\mathbf{X}|$ to a point in a higher-dimensional embedding space where bins that primarily contain energy from the same same source are near each other and bins that belong to different sources are far from each other.
Once trained, the network is used to embed a new magnitude spectrogram representing an auditory scene. Mask assignments are made by clustering time-frequency points in the higher dimensional embedding, assigning elements in the same cluster to the same source. With classical deep clustering, a clustering algorithm such as K-means is applied to the embedding space to determine the source assignments for every time-frequency point, which are then used to make a mask.

The Chimera architecture for audio source separation is a network architecture that combines deep clustering with a more traditional signal reconstruction loss \cite{erdogan2015phase}. In Chimera, there are two ``heads'' attached to a single ``body''. One head is trained with the deep clustering loss function while the other is trained to create the mask. Both heads are trained simultaneously during training. In Chimera, the mask inference head directly creates a mask that is element-wise multiplied to the mixture spectrogram to recover the sources. In this case, the deep clustering head acts as a regularizer for the mask inference head. The mask inference head is trained using a phase sensitive approximation (PSA)~\cite{erdogan2015phase} (or similar). 

\subsection{Adding a Transcription Head: Cerberus}
Since Chimera networks have been shown to outperform networks that do only deep clustering or only mask inference, we chose to use both heads in our work. We propose adding a third head to a Chimera network for further multitask learning. We call the proposed architecture a Cerberus network. The new head is used for the automatic transcription of musical mixtures containing multiple polyphonic instruments. All together, the output of the Cerberus architecture has three heads: a deep clustering head producing embeddings, a mask inference head creating masks that are applied to mixture spectrograms, and a transcription head that produces a piano roll transcription for each instrument in the mixture. The transcription estimate is a real-valued matrix with shape $T \times P \times I$, where $T$ is time frames (aligned with STFT time frames), $P$ is the number of possible pitches, and $I$ is the number of instruments to transcribe. The value of $I$ is fixed, corresponding to the number of sources. Once trained, the output of this head is quantized to binary to produce a piano roll transcription.


\begin{table}[t]
\small
\begin{center}
\begin{tabular}{ccc ? c ? c|c|c}

\multicolumn{3}{c ? }{} & Separation & \multicolumn{3}{c}{Transcription} \\
\multicolumn{3}{c ? }{Loss Type} & SDR (dB) & P & R & F1 \\ \thline \hline \hline
$\textrm{DC}_{1.0}$  &                     &                     &  8.5 & \cellcolor{black!25}  & \cellcolor{black!25}  & \cellcolor{black!25}  \\  \hline
                     & $\textrm{MI}_{1.0}$ &                     & \textbf{10.0} & \cellcolor{black!25}  & \cellcolor{black!25}  & \cellcolor{black!25}  \\ \hline
                     &                     & $\textrm{TR}_{1.0}$ & \cellcolor{black!25} &  0.48  & 0.43  & 0.44   \\ \thline
$\textrm{DC}_{0.5}$  & $\textrm{MI}_{0.5}$ &                     &  9.8 & \cellcolor{black!25}  & \cellcolor{black!25}  & \cellcolor{black!25}  \\ \hline
$\textrm{DC}_{0.2}$  &                     & $\textrm{TR}_{0.8}$ &  9.3 & 0.48  & 0.41  & 0.43  \\ \hline
                     & $\textrm{MI}_{0.2}$ & $\textrm{TR}_{0.8}$ &  9.8 & \textbf{0.51} & \textbf{0.46} & \textbf{ 0.47} \\ \thline
$\textrm{DC}_{0.1}$  & $\textrm{MI}_{0.1}$ & $\textrm{TR}_{0.8}$ & \textbf{10.0} & \textbf{0.51} & \textbf{0.45} & \textbf{ 0.47} \\  
\end{tabular}
\caption{Cerberus networks trained and tested on piano + guitar mixtures. Each row is a distinct network, trained with a distinct combination of three loss functions: Deep Clustering (DC), Mask Inference (MI) and Transcription (TR). The weight applied to a loss function is shown as a subscript. Evaluation measures for the transcription task are precision (P), recall (R) and F1. Evaluation for separation is scale-dependent source to distortion ratio (SDR). Higher values are better. The value in each cell is on the testing data, averaged across both instruments. Grey cells indicate the network was not trained for that task. We note that these results from note on/off precision, recall and F1 are competitive with state-of-the-art with models for isolated piano transcription.}
\label{tab:piano_guitar}
    
\end{center}
\vspace{-0.5cm}
\end{table}{}

The system is trained using a weighted linear combination of three loss functions. 

\vspace{-0.65cm}
\begin{align}
\mathcal{L}_{\text{Cerberus}} = \alpha \mathcal{L}_{\text{DC}} + \beta \mathcal{L}_{\text{MI}} + \gamma \mathcal{L}_{\text{TR}}
\label{eq:cerb_loss}
\end{align}{}

\vspace{-0.65cm}

The deep clustering \cite{hershey2016deep} loss ($\mathcal{L}_{\text{DC}}$) and the mask inference \cite{erdogan2015phase} loss ($\mathcal{L}_{\text{MI}}$) are those used in the Chimera network. For transcription loss $\mathcal{L}_{\text{TR}}$, we found that  $L_2$ distance to a MIDI-derived piano-roll score works best as the loss (see the next section for details). For inference, any of the three heads can be used, depending on the task.

\vspace{-0.5cm}
\section{Experimental Design}
Our experiments are designed to answer two questions. The first is whether learning to simultaneously separate and transcribe using a single network helps or hurts performance on either task, versus learning the tasks independently. The second is whether all three heads are needed for a effective separation and transcription system. 

\vspace{-0.25cm}
\subsection{Datasets and evaluation}
\vspace{-0.25cm}

To train a Cerberus network, a dataset is required that contains mixtures, isolated sources, and ground truth transcriptions for those sources. Slakh2100~\cite{manilow2019cutting} is one such dataset. It is comprised of 2100 mixtures made with sample-based professional synthesizers along with isolated sources and accompanying MIDI data for each source. We downsample the audio to 16 kHz. To make an example, we pick a mix, pick a subset of the sources (e.g. piano, guitar, bass) in the mixture, and then pick a 5 second segment where all the desired sources have at least 10 note onsets with MIDI velocity above $30$. The source audio for the desired sources is combined to make a mixture of just those sources. STFTs with 1024-point window size and 256 sample hop are calculated from mixture segments as inputs to the network. The musical score is the accompanying MIDI data binarized with a velocity threshold of $30$.

Using this procedure, we made 4 sets, each with 20000 segments (28 hours) for training, 3000 (4 hours) for validation, and 1000 (1.4 hours) for testing. The instrument combinations for the four sets were piano + guitar (set 1), piano + guitar + bass (set 2), and piano + guitar + bass + drums (set 3), and piano + guitar + bass + drums + strings (set 4).

In addition to the synthesized audio data, we evaluate on recordings of real instruments. To our knowledge, no large dataset exists that contains real-world recordings of mixtures, isolated sources, and ground truth transcriptions. Thus, we make mixtures from two datasets of real solo instrument recordings. The first is the MAPS\footnote{The MUS partition of both ENSTDkAm and ENSTDkCl.}~\cite{emiya2009multipitch}, which contains 30 live piano performances of classical music recorded with two microphones (60 clips total), and a Disklavier MIDI recorder. The second dataset is GuitarSet~\cite{xi2018guitarset}, which contains 360 30-second guitar excerpts in 5 styles. We downsample to 16 kHz, select 5 second segments with at least 10 note onsets, and use the same STFT parameters. We randomly selected segments from each dataset to make 1000 instantaneous mixtures with accompanying sources and score data. These mixes are incoherent, meaning the mixed segments come from different songs with different tempos and key signatures (MAPS is classical piano, GuitarSet is rock guitar). This data is highly dissimilar to the network's training data.

For source separation, we use the scale-dependent source-to-distortion ratio~\cite{le2019sdr} for evaluation. For transcription we use precision, recall, and F1-score of note onsets and offsets using the \textit{mir\_eval} toolbox \cite{raffel2014mir_eval}. These are both commonly used measures in the literature for their respective tasks.

\vspace{-0.25cm}
\subsection{Networks we evaluate}
\vspace{-0.25cm}

All the networks we trained use a stack of 4 bidirectional long short-term memory (BLSTM) layers. Each BLSTM has 300 hidden units. We trained each network for 100 epochs using an Adam optimizer with an initial learning rate of 2e-4, a batch size of $40$, and a sequence length of $400$ frames. Each network had three heads. The first head maps each time-frequency point to a 20-dimensional embedding space, with sigmoid activation and unit-normalization. The second head outputs masks for each of the sources we trained the network to separate (between 2 and 5 masks), with a softmax activation across the masks. The third head outputs transcriptions for each source and has a sigmoid activation. Each transcription contains $88$ pitches and when each pitch is active. For evaluation, we binarized the network's transcriptions using an experimentally determined threshold of $0.8$, except for drums, which was set to $0.1$. Each network was initialized with the same set of weights. The only difference between the networks is the training data (for which instrument combination to separate) and the weights on the the loss functions: deep clustering (DC), mask inference (MI), and transcription (TR). 

\vspace{-0.25cm}
\section{Results}
For the first set of experiments, we trained a Cerberus network to separate and transcribe mixtures of one piano and one guitar from Slakh2100. The dataset for this experiment include acoustic and electric pianos and acoustic, electric, and distorted guitars. In this experiment set, we set certain loss weights to zero to make seven combinations of Cerberus networks. Turning off the transcription loss ( \textit{i.e.,} $\gamma=0.0$ in Equation \ref{eq:cerb_loss}) results in a standard Chimera network. In the Chimera network (4th row of Table \ref{tab:piano_guitar}), we weighted the two separation losses equally. In the Cerberus and Chimera transcription networks, we observed that the scale of the transcription loss was much smaller than the scale of the separation losses. To counteract this, we more heavily weighted the transcription loss during training for these networks, while keeping the two separation losses at equal weight. We note that the values that we present are competitive with the state-of-the-art models for isolated piano transcription and that transcription F1-scores for an untrained network are on the order of 1e-3.

\begin{table}[]
\begin{center}
\footnotesize

\begin{tabular}{ l ? l ? c ? c|c|c}
Test & & \multicolumn{1}{c ? }{Separation} & \multicolumn{3}{c}{Transcription} \\
Dataset & {\small Network Type} & SDR (dB) & P & R & F1 \\ \tthline \hline \hline

\multicolumn{1}{r ?}{GS} & Transcription Only  &  \cellcolor{black!25} & 0.08 & 0.08 & 0.08 \\ \hline
\multicolumn{1}{r ?}{GS} & \textbf{Cerberus} & \cellcolor{black!25} & \textbf{0.13} & \textbf{0.11} & \textbf{0.12} \\  \thline
M      & Transcription Only  &  \cellcolor{black!25} &  0.19  & 0.08  & 0.11 \\ \hline
M      & \textbf{Cerberus} & \cellcolor{black!25} & \textbf{0.19} &\textbf{ 0.10} & \textbf{0.12} \\ \thline
M + GS & Deep Clustering Only & 4.3 & \cellcolor{black!25}  & \cellcolor{black!25}  & \cellcolor{black!25}  \\ \hline
M + GS & Mask Inference Only & 4.1 & \cellcolor{black!25}  & \cellcolor{black!25}  & \cellcolor{black!25}  \\ \hline
M + GS & Chimera  & 4.5  & \cellcolor{black!25}  & \cellcolor{black!25}  & \cellcolor{black!25}  \\ \hline
M + GS & Transcription Only &  \cellcolor{black!25} & 0.14 & 0.08 & 0.09 \\ \hline
M + GS & \textbf{Cerberus} & \textbf{5.0} & \textbf{0.16} & \textbf{0.10} & \textbf{0.12} \\ 
\end{tabular}
\caption{Piano/Guitar performance on MAPS and GuitarSet data using networks from Table \ref{tab:piano_guitar} trained on Slakh2100. \textit{M} means MAPS recordings in isolation, \textit{GS} means GuitarSet recordings in isolation, and \textit{M+GS} means incoherent mixtures of recordings from MAPS and GuitarSet. Grey cells indicate the network was not trained for that task. Evaluation measures for the transcription task are precision (P), recall (R) and F1. Evaluation for separation is scale-dependent source to distortion ratio (SDR). Higher values are better.  }
\label{tab:maps}
\end{center}
\vspace{-0.5cm}
\end{table}

The results in Table \ref{tab:piano_guitar} suggest that transcription and separation can be learned jointly, given the correct training regime. First, we find that the best performing model for both transcription and separation was the Cerberus model, which surpassed or tied the highest SDR and precision, recall, and F1 scores of the remaining models. Combining the mask inference and transcription objectives resulted in higher transcription performance but lowered separation performance very slightly. Finally, combining the deep clustering and transcription objectives resulted in a large jump in SDR over just deep clustering, suggesting some natural synergy between the two tasks.

Next, we took networks trained on the synthesized Slakh dataset and evaluated them on the real-world dataset we generated from MAPS and GuitarSet. The results are shown in Table \ref{tab:maps}. First, we notice a significant drop in separation and transcription performance. This is, in large part, due to the major differences between the training and test data, namely that the training data contains all coherent mixes (instruments all playing the same song) and the test data is all incoherent mixes (instruments mixed from different songs). We note that the Cerberus model that was trained with all three loss functions out-performs all of the single-task networks for both separation and transcription, suggesting that our multi-task approach leads to better generalization.

\begin{table}[t]
\begin{center}
\small
\begin{tabular}{c  l ? c ? c|c|c}

\multicolumn{2}{l ?}{Cerberus trained/tested } & Separation & \multicolumn{3}{c}{Transcription} \\ 
\multicolumn{2}{l ?}{on data that contains:} & SDR (dB) & P & R & F1 \\ \thline \thline \hline \hline

\multirow{3}{*}{3 Sources} & \makebox[0.1cm]{} Piano   &  7.6   & 0.44  & 0.42 & 0.42  \\ 
& \makebox[0.1cm]{{\scriptsize \& }} Guitar  &  6.9  & 0.46  & 0.35  & 0.38  \\
& \makebox[0.1cm]{{\scriptsize \& }} Bass &  10.1  & 0.85  & 0.80  & 0.82  \\ \thline  \hline \hline 
\multirow{4}{*}{4 Sources}& \makebox[0.1cm]{} Piano  & 6.1 & 0.38  & 0.36  & 0.36  \\ 
& \makebox[0.1cm]{{\scriptsize \& }} Guitar   &  5.8  & 0.42  & 0.32  & 0.34  \\
& \makebox[0.1cm]{{\scriptsize \& }} Bass  &   7.7  &  0.82 & 0.78  & 0.79  \\ 
& \makebox[0.1cm]{{\scriptsize \& }} Drums* &   11.3  & 0.61  & 0.76  & 0.63  \\ \thline \hline \hline 
\multirow{5}{*}{5 Sources}& \makebox[0.1cm]{} Piano  &  3.4   & 0.31  & 0.28  &  0.28 \\ 
& \makebox[0.1cm]{{\scriptsize \& }} Guitar  &  3.1  & 0.29  & 0.20  & 0.22  \\
& \makebox[0.1cm]{{\scriptsize \& }} Bass  &  6.4  & 0.77  & 0.72  &  0.74 \\ 
& \makebox[0.1cm]{{\scriptsize \& }} Drums* &  10.6  & 0.62  &  0.75  & 0.64  \\ 
& \makebox[0.1cm]{{\scriptsize \& }} Strings &  4.1  & 0.39  & 0.35  &  0.35 \\ 
\end{tabular}
\caption{Results for individual instruments from three Cerberus networks trained on different sets of instrument combinations, separated by horizontal lines. Each model has its own training, validation, and test set which depend on the instruments it is trained to separate and transcribe. Drum (*) transcription evaluation measures note onset, all other instruments are note on/off precision/recall/f-score.}
\label{tab:instruments}
\end{center}
\vspace{-0.5cm}
\end{table}

Finally, we trained and tested a Cerberus model on data sets of increasing numbers of simultaneous polyphonic instruments to see how the system scales up to more complex mixtures. The results, shown in Table \ref{tab:instruments}, show that as we add more sources to the mixture, performance across both separation and transcription predictably degrades. While piano, guitar, and strings results are low in the most complex setup (bottom rows), bass and drums can still be separated and transcribed from complex mixtures.


\vspace{-0.5cm}
\section{Conclusion}
\vspace{-0.4cm}

We introduced an architecture to simultaneously transcribe and separate multiple instruments in a musical mixture. This architecture, called Cerberus, has three ``heads'': one for transcription, one for deep clustering, and one for mask inference. Cerberus networks are more effective at both tasks than single-task networks on both real and synthesized data. Future work could include more involved network architectures, dedicated losses for note onsets and velocities\cite{hawthorne2017onsets}, and training on existing score-aligned recordings isolated instruments to strengthen transcription on real recordings, as well as training on real multi-instrument recordings when aligned score data becomes available.

\bibliographystyle{IEEEbib}
\bibliography{strings,refs}

\end{document}